\documentclass[aps,preprint,nofootinbib]{revtex4}
\begin{document}
\title{\Large \bf A Slowly Rotating Charged Black Hole in Five Dimensions}
\author{\large A. N. Aliev}
\email{aliev@gursey.gov.tr}
\affiliation{ Feza G\"ursey Institute,
P.K. 6  \c Cengelk\" oy, 81220 Istanbul, Turkey}

\begin{abstract}

Black hole solutions in higher dimensional Einstein and
Einstein-Maxwell gravity  have been discussed by Tangherlini as well as
Myers and Perry a long time ago. These solutions are the
generalizations of the familiar Schwarzschild, Reissner-Nordstrom
and Kerr solutions of four-dimensional general relativity. However,
higher dimensional generalization of the Kerr-Newman
solution in four dimensions has not been found yet. As a first step
in this direction I shall report on a new solution of the
Einstein-Maxwell system of equations that describes an electrically
charged and slowly rotating black hole in five dimensions.

\end{abstract}
\maketitle
\section{Introduction}

The study of black holes in higher dimensions is motivated by
several reasons. First of all, it is related to the intrinsic
properties of human nature; one knows the fundamental features of
black holes in four dimensions, such as the equilibrium and
uniqueness properties, astonishing thermodynamical properties,
Hawking's effect of evaporation of mini black holes and wonders what
will happen when going over into higher dimensions. Strong
motivation comes from developments in string/M-theory, which is
believed to be the most consistent approach to quantum theory of
gravity in higher dimensions. Black holes may have a crucial role in
the analysis of dynamics in higher dimensions as well as in the
compactification mechanisms. In particular, to test novel
predictions of string/M-theory microscopic black holes may serve as
good theoretical laboratories. As an example, one can recall the
statistical-mechanical calculation of the Bekenstein-Hawking entropy
for a class of supersymmetric black holes in five dimensions which
is thought of as one of the remarkable results in string
theory  \cite{sv, bmpv}.

Another strong motivation for the interest in higher dimensional
black holes originates from the brane-world gravity theories, which
predict a new fundamental scale of quantum gravity being of the
order of TeV-scale \cite{ADD}-\cite{RS2}. One of the exciting
signatures of these models is the possibility of TeV-size mini black
hole production at future colliders \cite{dl}. Furthermore, it has
been argued that if the radius of the event horizon is much smaller
than the size-scale of extra dimensions $(\,r_{+} \ll L \,)$, these
black holes to a good enough approximation can be described by the
classical solutions of higher dimensional Einstein's field
equations. In this framework the most interesting black hole
solutions are the Tangherlini solution \cite{tang} for a static
black hole and the Myers-Perry solution \cite{mp}  for a stationary
black hole in higher dimensions. In recent developments the
different physical properties of these solutions have been discussed
in a number of papers \cite{gibbons1}- \cite{af} (see also Refs.
\cite{marco}, \cite{kanti} for reviews). Furthermore, it has been
shown that the Myers-Perry solution is not unique in five
dimensions; there exists a  rotating black ring solution with the
horizon topology of $\, S^2\times S^1\,$ which could have the same
mass and spin as the Myers-Perry solution \cite{er}.

It is clear that black holes produced at colliders may in general
have an electric charge as well as other type of charges. Therefore
the study of charged black hole solutions in higher dimensions
becomes of great importance. The first black hole solution to the
higher dimensional Einstein-Maxwell equations was found by
Tangherlini \cite{tang}. This solution is the generalization of
the familiar Reissner-Nordstrom solution for a static and
electrically charged black hole in ordinary general relativity.
However, the case of charged rotating black holes has been basically
discussed within certain supergravity theories \cite{town}, as well as
in the context of string theory (see review papers \cite{youm}-\cite{peet}).
As for the counterpart of the Kerr-Newman
solution in higher dimensions, that is the charged generalization of
the Myers-Perry solution in the Einstein-Maxwell gravity,
it still remains to be found. Here I shall discuss a
new solution of the Einstein-Maxwell field equations describing an
electrically charged and slowly rotating black hole in five
dimensions.

\section{Rotating black hole with electric charge}

As is known the Kerr metric for a stationary black hole in four
dimensions is uniquely characterized  by an axis of rotation
consistent with the fact that there exists only one independent
Casimir invariant of the rotation group $\; SO(3)\;$. However, in
five dimensions the rotation group is $\; SO(4)\;$ which posseses
two independent Casimir invariants. These two Casimir invariants are
associated with two independent rotations of the system. Thus, a
rotating black hole in five dimensions may have two distinct planes
of rotation specified by appropriate azimuthal coordinates, rather
than an axis of rotation. In accordance with this, the stationary
and asymptotically flat Myers-Perry metric admits three commuting
Killing vector fields
\begin{equation}
{\bf \xi}_{(t)}= \partial / \partial t\,, ~~~~ {\bf \xi}_{(\phi)}=
\partial / \partial \phi \, , ~~~~ {\bf \xi}_{(\psi)}= \partial /
\partial \psi \,, \label{killing}
\end{equation}
which reflect the time-translation invariance and bi-azimuthal
symmetry of the metric in five dimensions. The explicit form of the
metric written in Boyer-Lindquist type coordinates is given by
\begin{eqnarray}
ds^2 & = & -dt^2 +\Sigma\,\left(\frac{r^2}{\Pi}\,dr^2 + d\theta^2
\right) +(r^2 + a^2)\,\sin^2\theta \,d\phi^2
+(r^2 + b^2)\,\cos^2\theta \,d\psi^2 \nonumber \\
&  & +\,\frac{m}{\Sigma}\, \left(dt - a\, \sin^2\theta \,d\phi -
b\,\cos^2\theta \,d\psi \right)^2\,\,, \label{metric}
\end{eqnarray}
where the functions $\,\Sigma\,$ and $\,\Pi\,$ are given as
\begin{equation}
\Sigma=r^2+a^2 \,\cos^2\theta + b^2 \,\sin^2 \theta, \;\;\;\;\;\;
\Pi= (r^2 + a^2)(r^2 + b^2) -m \, r^2 \,,
\end{equation}
$\,m \,$  is a parameter related to the physical mass of the
black hole, the parameters  $\,a \,$ and $\,b \,$ are associated
with its two independent angular momenta .\footnote{The expression for
$\,\Sigma\,$ in \cite{af} has a misprint. The parameters $\,a \,$ and $\,b\,$ must be interchanged.}

Earlier in \cite{af} we studied the electromagnetic properties of the
Myers-Perry black assuming that it may possess a small enough electric charge,
such that the spacetime is still well described by the unperturbed metric (\ref{metric}) .
It is remarkable that in this case the five-vector potential for the source-free Maxwell
field can be constructed using the time-translation invariance
property of the metric . Indeed, the Maxwell equations
for the vector potential in the Lorentz gauge
\begin{equation}
A^{\mu}_{\;;\;\mu} =0\,, \label{lorentz}
\end{equation}
have the form
\begin{equation}
{A^{\mu\,;\,\nu}}_{\,;\,\nu}- {R^{\mu}}_{\nu}\,A^\nu=0\,\,.
\label{max}
\end{equation}
On the other hand, any Killing vector $\, {\bf \xi}\,$ satisfies the
equation
\begin{equation}
{\xi^{\mu\,;\,\nu}}_{\,;\,\nu} + {R^{\mu}}_{\nu}\,\xi^\nu=0\,\,.
\label{kileq}
\end{equation}
Comparing these two equations  we see
that in a vacuum spacetime $\,({R^{\mu}}_{\nu}=0)\,$, the Killing
vector can serve as a vector potential for a test Maxwell field
\cite{papa}. Applying this fact to our case we can consider the $5$-vector potential of
the form
\begin{equation}
A^{\mu} = \alpha \,\, \xi^{\mu}_{(t)}
\label{potform}
\end{equation}
where $\,\alpha\,$ is an arbitrary parameter which can be fixed from
examining the Gauss integral for the electric charge of the black
hole
\begin{equation}
Q=\frac{1}{4\,\pi^2}\,\oint \,F^{\mu\,\nu}\, d\,^3 \Sigma_{\mu\;\nu}
\,\,. \label{charge}
\end{equation}
The integral is taken over the three-sphere at spatial
infinity
\begin{equation}
d\,^3 \Sigma_{\mu\;\nu}= \frac{1}{3!}\,\sqrt{-g}\,
\epsilon_{\mu\,\nu\,\alpha\,\beta\, \gamma} \,d\,x^{\alpha} \wedge
d\,x^{\beta} \wedge d\,x^{\gamma}\,\,. \label{3sphere}
\end{equation}
Taking this into account and requiring the vanishing behaviour of
the potential at infinity we finally obtain
\begin{equation}
A = -\frac{Q}{2\,\Sigma}\, \left(d\,t - a\,\sin^2\theta\,d\,\phi -
b\,\cos^2\theta\,d\,\psi\right)\,\,. \label{potform1}
\end{equation}
Accordingly, the associated electromagnetic $2$-form field can be
written as
\begin{eqnarray}
F&=&\,\frac{Q}{\Sigma\,^2}\,\left(r\,d\,r \wedge d\,t
+(b^2-a^2)\,\sin\theta \cos\theta \,d\,\theta \wedge d\,t \right)
 \nonumber\\[3mm] & &
- \frac{Q\,a\,\sin\theta}{\Sigma\,^2}\,\left(r\,\sin\theta\,
d\,r\wedge d\,\phi - (r^2+a^2)\,\cos\theta\,d\,\theta  \wedge
d\,\phi \right)
 \nonumber\\ [3mm]   & &
- \frac{Q\,b\,\cos\theta}{\Sigma\,^2}\,\left(r\,\cos\theta\,
d\,r\wedge d\,\psi + (r^2+b^2)\,\sin\theta\,d\,\theta  \wedge
d\,\psi \right) \,\,, \label{eft}
\end{eqnarray}
while the contravariant components of the electromagnetic field
tensor are given by
\begin{eqnarray}
F^{tr}&=&\frac{Q\, (r^2+a^2)(r^2+b^2)}{r\,\Sigma\,^3}\,\,, ~~~~~
F^{t\theta}= \frac{Q\,(b^2-a^2)\,\sin2\theta}{2\,\Sigma\,^3}\,\,,
\nonumber \\[3mm]
F^{r\phi}& =& - \frac{Q\,a\,(r^2+b^2)}{r\,\Sigma\,^3}\,\,,
~~~~~~~~~~~~~~~ F^{r\psi} = -
\frac{Q\,b\,(r^2+a^2)}{r\,\Sigma\,^3}\,\,,
 \nonumber \\[3mm]
F^{\theta\phi}& =&\frac{Q\,a\,\cot\theta}{\Sigma\,^3}\,\,,
~~~~~~~~~~~~~~~~~ F^{\theta\psi} = -
\frac{Q\,b\,\tan\theta}{\Sigma\,^3}\,\,. \label{emtcontra}
\end{eqnarray}

Next, we suppose that the electric charge is no longer small
that its electromagnetic field does influence the
metric of space-time around the black hole.

\section{Metric ansatz}

It is clear that for an arbitrary electric charge of the black hole
one must consistently solve the system of the
Einstein-Maxwell equations
\begin{equation}
{R^{\mu}}_{\nu}= 8 \pi G {M^{\mu}}_{\nu}\,\,,
\label{einstein}
\end{equation}
\begin{equation}
\partial_{\nu} (\sqrt{-g}\,F^{\mu \nu})= 0 \,\,, \label{max1}
\end{equation}
where the source-term on the right-hand-side of equation
(\ref{einstein}) is given by
\begin{equation}
{M^{\mu}}_{\nu}={T^{\mu}}_{\nu}-\frac{1}{3}\,{\delta^{\mu}}_{\nu}\,T
\,.
\end{equation}
Substituting into this expression the explicit form of the energy-momentum tensor
 for the electromagnetic field
\begin{equation}
{T^{\mu}}_{\nu}=  \frac{1}{2\,\pi^2}\left(F^{\mu \alpha} F_{\nu
\alpha}-\frac{1}{4}\,{\delta^{\mu}}_{\nu}\,F_{\alpha \beta} F^{\alpha \beta}  \right) \,
\label{emt}
\end{equation}
and its trace $\,T\,$ we put it in the form
\begin{equation}
{M^{\mu}}_{\nu}= \frac{1}{2\,\pi^2}\left(F^{\mu \alpha} F_{\nu
\alpha}-\frac{1}{6}\,{\delta^{\mu}}_{\nu}\,F_{\alpha \beta} F^{\alpha \beta}\right) \,.
\label{sterm1}
\end{equation}

For the sake of simplicity, let us suppose that the five-dimensional
black hole possesses only one angular momentum (i. e. $\,b=0\,$) and
consider an ansatz for the stationary metric in the Boyer-Lindquist
type form, that is involving the only off-diagonal component with
indices $\,t\,$ and $\,\phi\,$. Namely, we start with the metric
ansatz
\begin{eqnarray}
ds^2 & = & -\left(1-H \right)\,dt^2 +\frac{\Sigma}{\Delta}\,\,dr^2 +
\Sigma \,d\theta^2 - 2\, a H \sin^2\theta\,d
t\,d\phi \nonumber \\[2mm]
&  & +\,\left(r^2 + a^2 + H a^2\,\sin^2\theta
\right)\sin^2\theta\,d\phi^2 + r^2\cos^2\theta \,d\psi^2 \,\,,
\label{cmetric}
\end{eqnarray}
where we have introduced an arbitrary scalar function $\,H\,$
depending only on the coordinates $\,r\,$ and $\,\theta\,$ and the
function $\,\Delta\,$ is given by
\begin{equation}
\Delta=r^2+a^2 -H \Sigma\,\,. \label{newd}
\end{equation}
Performing straightforward calculations it is easy to show that the
five-vector potential (\ref{potform1}) with $\,b=0\,$ satisfies the
Maxwell equation (\ref{max1}) in the metric (\ref{cmetric})
independently on the explicit form of the function $\,H\,$. Thus,
one can use the relations (\ref{eft}) and (\ref{emtcontra}) to
calculate the non-vanishing components of the source tensor
(\ref{sterm1}). We find that
\begin{eqnarray}
{M^t}_{t}&=&\frac{Q^2}{6\pi^2 \Sigma^{\,4}}\,\left(2 \Sigma -4 r^2
-3 a^2\right)\,\,, ~~~~~~  {M^r}_{r}= -\frac{Q^2}{6\pi^2
\Sigma^{\,4}}\,\left(\Sigma + r^2\right)\,\,,
\nonumber \\[3mm]
{M^\theta}_{\theta}& =& \frac{Q^2}{6\pi^2 \Sigma^{\,4}}\,\left(2
\Sigma - r^2\right) \,\,, ~~~~~~~~~~~~~~~
{M^\phi}_{\phi}=-\frac{Q^2}{6\pi^2 \Sigma^{\,4}}\,\left(\Sigma -2\,
r^2 -3 a^2\right)\,\,,\nonumber \\[3mm]
{M^\psi}_{\psi}&=& -\frac{Q^2}{6\pi^2 \Sigma^{\,4}}\,\left(\Sigma -
2 r^2\right)\,\,, ~~~~~~~~~~~~
{M^\phi}_{t}=-\frac{Q^2 a}{2\pi^2\Sigma^{\,4}}=-\frac{{M\,^t}_{\phi}}{(r^2+a^2)\sin^2\theta}\,\,.
 \label{sterm2}
\end{eqnarray}
One can easily check that  there exists the relation
\begin{equation}
{M^{t}}_{t}+{M^{\phi}}_{\phi}+{M^{\psi}}_{\psi}=0 \label{soeq}
\end{equation}
between the components of the source tensor.

On the other hand, the non-vanishing components of the Ricci tensor calculated for the metric
(\ref{cmetric}) have the form
\begin{eqnarray}
{R^t}_{t}&=&\frac{r^2+a^2}{2 \Sigma}\,\left(H_{rr}+\frac{H_{\theta\theta}}{\Delta} +\frac{\Sigma}{\Delta^2}\,H_{\theta}^2
+\frac{3 r^2+a^2}{r^2+a^2}\,\frac{H_{r}}{r} \right) -\frac{a^2 H^2}{\Delta \Sigma}\left(2 \cos2\theta-
 \frac{a^2 H \sin^2 2\theta}{2 \Delta}\right)
\nonumber \\[3mm]
& &
+\frac{H_{\theta}}{\Delta \Sigma} \left[(r^2+a^2) \cot 2\theta -a^2 H \sin2 \theta \left(\frac{1}{2} +\frac{r^2+a^2}{\Delta}\right)\right]\,\,,
\label{ricci00}
\end{eqnarray}
\begin{eqnarray}
{R^r}_{r}&=&\frac{1}{2}\,\left(H_{rr}-\frac{H_{\theta\theta}}{\Delta} -\frac{\Sigma}{\Delta^2}\,H_{\theta}^2
+\frac{\Sigma + 2 r^2}{\Sigma}\,\frac{H_{r}}{r} \right) -\frac{H_{\theta}}{\Delta}\left( \cot2\theta-
 \frac{r^2+a^2}{\Delta \Sigma}\,a^2 \sin 2\theta \right)
\nonumber \\[3mm]
& &
+\frac{a^2 H}{\Delta^2 \Sigma} \left[(r^2+a^2)\left(1+3 \cos 2\theta + \frac{a^2 \sin^2 2 \theta}{2 \Sigma}\right) +2 H^2\Sigma \cos^2 \theta\right]
\nonumber \\[3mm]
& &
-\frac{2\, a^2 H^2}{\Delta^2}\left(1+2 \cos 2\theta + \frac{3 a^2 \sin^2 2 \theta}{4 \Sigma}\right)\,\,,
\label{ricci11}
\end{eqnarray}
\begin{eqnarray}
{R^\theta}_{\theta}&=&\frac{r}{\Sigma}\,\,H_{r}-\frac{H_{\theta}}{2 \Delta^2}
\left(\Sigma H_\theta-2 a^2 H \sin2\theta \right) +\frac{2 r^2 H}{\Delta^2 \Sigma^2} \left[(r^2+a^2)^2+ \Sigma^2 H^2 \right]
\nonumber \\[3mm]
& & -\frac{H^2}{2 \Delta^2 \Sigma} \left[8 r^2 (r^2+a^2)-a^4 \sin^2
2\theta \right]\,\,, \label{ricci22}
\end{eqnarray}
\begin{eqnarray}
{R^\phi}_{\phi}&=&-\frac{a^2 \sin^2 \theta}{2 \Sigma}\,\left(H_{rr}+\frac{H_{\theta\theta}}{\Delta} +\frac{\Sigma}{\Delta^2}\,H_{\theta}^2\right)
-\frac{2 r^2-a^2 \sin^2\theta}{2 \Sigma}\,\frac{H_{r}}{r} -\frac{a^2 H^2 \sin2\theta}{2 \Delta^2} H_\theta
\nonumber \\[3mm]
& &
-\frac{a^2 H_{\theta} \tan\theta}{2\Delta^2}\left[\frac{r^2+a^2}{\Sigma}\,\left(2+3 \cos2\theta\right)- H
\left(3+4 \cos 2\theta + \frac{3 a^2 \sin^2 2 \theta}{2 \Sigma}\right)\right]
\nonumber \\[3mm]
& & -\frac{2 H^2}{\Delta^2 \Sigma}\left[a^4 \sin^2\theta
\left(1+\cos^2\theta\right)+ 2 r^2 \left(r^2 +a^2 +a^2
\sin^2\theta\right)\right]
\nonumber \\[3mm]
& & +\frac{2 H (r^2+a^2)}{\Delta^2 \Sigma} \left(r^2+2 a^2
\sin^2\theta +r^2 H^2 \right)\,\,, \label{ricci33}
\end{eqnarray}
\begin{eqnarray}
{R^\phi}_{t}&=&\frac{a}{2 \Sigma}\,\left[\left(H_{rr}
+\frac{H_{r}}{r}
+\frac{H_{\theta\theta}}{\Delta}+\frac{\Sigma}{\Delta^2}\,H_{\theta}^2\right)
-\frac{2  H}{\Delta^2} \left(r^2 +a^2 +r^2 H^2 -H(a^2+2
r^2)\right)\right]
\nonumber \\[3mm]
& & +\frac{a H_\theta}{\Delta^2\sin 2
\theta}\left[\frac{r^2+a^2}{\Sigma}\,\left(1+2 \cos2\theta\right)- H
\left(2+3 \cos 2\theta + \frac{3 a^2 \sin^2 2 \theta}{2
\Sigma}\right)\right]
\nonumber \\[3mm]
& & + \frac{a H^2 \cot\theta}{\Delta^2}\,\,\,,
 \label{ricci30}
\end{eqnarray}
\begin{eqnarray}
{R^t}_{\phi}&=&-(r^2 +a^2)\sin^2\theta {R^\phi}_{t}+ \frac{a H \sin
2\theta}{2\Delta} \left(\frac{a^2 H
\sin2\theta}{\Sigma}-H_\theta\right)\,\,, \label{ricci03}
\end{eqnarray}
\begin{eqnarray}
{R^r}_{\theta}&=& \Delta {R^\theta}_{r}=\frac{1}{2}
\left(H_{r\theta}+\frac{H_{\theta}}{r}\right)- \frac{a^2
\sin2\theta}{2 \Sigma} \left(H_r+\frac{\Sigma-2 r^2}{r\,\Sigma}\,H
\right)\,\,, \label{ricci12}
\end{eqnarray}
\begin{eqnarray}
{R^\psi}_{\psi}&=&\frac{1}{r \Sigma}\, \frac{\partial}{\partial r}\, \left(\Sigma\, H\right)\,\,,
\label{ricci44}
\end{eqnarray}
where the indices $r$ and $\theta$ at the scalar function $H$ denote
its differentiation with respect to the coordinates $r$ and $
\theta$. Substituting the quantities (\ref{sterm2}) and
(\ref{ricci00})-(\ref{ricci44}) into Einstein's equation
(\ref{einstein}) we see that one can easily solve the equation with
$\, {R^\psi}_{\psi}\,$ and $\, {M^\psi}_{\psi}\,$, however the
solution does not satisfy the remaining set of equations. Thus,
there is no scalar function $\,H\,$ fulfilling both equations
(\ref{einstein})and (\ref{max1}) simultaneously. Earlier this fact
was also noted in \cite{mp} within the Kerr-Schild type metric
ansatz.

\section{Slow rotation}

Let us now consider the case  when the rotation of the black hole
occurs slowly enough that we can expand all the expressions above in
powers of the rotation parameter restricting ourselves only to the
linear order in $ \,a\,$ terms. Then we arrive at the following set of
the field equations
$$H_{rr}+\frac{3 H_r}{r} +\frac{H_{\theta\theta}}{\Delta}
+\frac{r^2}{\Delta^2}\,H_{\theta}^2 +\frac{2 \cot 2\theta}{\Delta}
H_\theta = -\frac{16 G}{3 \pi}\,\frac{Q^2}{r^6}\,\,,$$

$$H_{rr}+\frac{3 H_r}{r} -\frac{H_{\theta\theta}}{\Delta}
-\frac{r^2}{\Delta^2}\,H_{\theta}^2 -\frac{2 \cot 2\theta}{\Delta}
H_\theta = -\frac{16 G}{3 \pi}\,\frac{Q^2}{r^6}\,\,, $$

$$
H_{r} -\frac{r^3}{2 \Delta^2}\left[H_{\theta}^2- 4 H (1-H)^2\right]=
\frac{4 G}{3 \pi}\,\frac{Q^2}{r^5}\,\,,
$$

$$H_{r} +\frac{2 r^3 H}{\Delta^2}\left(1-H\right)^2 =
\frac{4 G}{3 \pi}\,\frac{Q^2}{r^5}\,\,,
$$
\begin{eqnarray}
H_{rr}+\frac{H_r}{r} +\frac{H_{\theta\theta}}{\Delta}
+\frac{r^2}{\Delta^2}\,H_{\theta}^2 -\frac{4 r^2 H}{\Delta^2}
\left(1-H\right)^2
\nonumber \\[3mm]
+\,\frac{2 r^2 (1-H)}{\Delta^2 \sin2\theta}\,H_\theta \left(1+2 \cos
2\theta-2 H \cos^2 \theta\right) &=& -\frac{8 G}{
\pi}\,\frac{Q^2}{r^6}\,\,, \label{sricci30}
\end{eqnarray}

$$H_{r} +\frac{2}{r} H = \frac{4 G}{3 \pi}\,\frac{Q^2}{r^5}\,\,,$$

$$H_{r\theta}+\frac{H_{\theta}}{r} =0\,\,.$$

The solution to these equations has the simple form
\begin{eqnarray}
H= \frac{m}{r^2} - \frac{q^2}{r^4}\,\,,
\label{fsol}
\end{eqnarray}
where in addition to the mass parameter, we have also introduced the charge parameter
\begin{eqnarray}
q=\pm \sqrt{\frac{2 G}{3 \pi}}\,\,Q\,\,\,.
\label{chargep}
\end{eqnarray}

Finally, substituting the expression (\ref{fsol}) into equations (\ref{cmetric}) and (\ref{newd}),
up to terms linear in $ \,a\,$, we obtain the metric for the charged and slowly rotating black hole in five dimensions
\begin{eqnarray}
ds^2 & = & -\left(1-\frac{m}{r^2}+\frac{q^2}{r^4} \right)\,dt^2
+\left(1-\frac{m}{r^2}+\frac{q^2}{r^4} \right)^{-1}\,\,dr^2 + r^2\left(
\,d\theta^2
 + \sin^2\theta\,d\phi^2 + \cos^2\theta \,d\psi^2\right)
\nonumber \\[2mm]
& & - \frac{2\, a \,\sin^2\theta}{r^2}\left(m-\frac{q^2}{r^2}\right)\,d t\,d\phi\,\,,
 \label{fmetric}
\end{eqnarray}
while the associated electromagnetic field is given by the $2$-form
\begin{eqnarray}
F&=&\,\frac{Q}{r^3}\,\left[\,d\,r \wedge d\,t
- a\,\sin\theta \left(\sin\theta\,
d\,r- r\,\cos\theta\,d\,\theta \right) \wedge d\,\phi \,\right] \,\,. \label{feft1}
\end{eqnarray}
This metric generalizes the Schwarzschild-Tangherlini solution to
the case of a five-dimensional slowly rotating black black hole with
one angular momentum. In the similar way, one can also write down a
corresponding five-dimensional solution with two independent angular
momenta. This will be done in a future work.

\end{document}